\documentclass[aip,apl, floatfix, reprint]{revtex4-1}

\usepackage{graphicx}
\usepackage{dcolumn}
\usepackage{bm}

\usepackage[utf8]{inputenc}
\usepackage[T1]{fontenc}
\usepackage{mathptmx}
\usepackage{etoolbox}
\usepackage{wrapfig}
\usepackage{float}

\makeatletter
\def\@email#1#2{%
 \endgroup
 \patchcmd{\titleblock@produce}
  {\frontmatter@RRAPformat}
  {\frontmatter@RRAPformat{\produce@RRAP{*#1\href{mailto:#2}{#2}}}\frontmatter@RRAPformat}
  {}{}
}%
\makeatother
\begin{document}

\preprint{AIP/123-QED}

\title[]{Wafer-Scale Fabrication of InGaP-on-Insulator for Nonlinear and Quantum Photonic Applications}
\author{Lillian Thiel}
 \email{lthiel@ucsb.edu, moody@ucsb.edu}
\affiliation{Electrical and Computer Engineering Department, University of California, Santa Barbara, CA 93106}

\author{Joshua E. Castro}
\affiliation{Electrical and Computer Engineering Department, University of California, Santa Barbara, CA 93106}

\author{Trevor J. Steiner}
\affiliation{Materials Department, University of California, Santa Barbara, CA 93106}

\author{Catherine L. Nguyen}
\affiliation{Thorlabs Crystalline Solutions, Santa Barbara, California 93101}

\author{Audrey Pechilis}
\affiliation{Physics Department, University of California, Santa Barbara, CA 93106}

\author{Liao Duan}
\affiliation{Physics Department, University of California, Santa Barbara, CA 93106}

\author{Nicholas Lewis}
\affiliation{Electrical and Computer Engineering Department, University of California, Santa Barbara, CA 93106}

\author{Garrett D. Cole}
\affiliation{Thorlabs Crystalline Solutions, Santa Barbara, California 93101}

\author{John E. Bowers}
\affiliation{Electrical and Computer Engineering Department, University of California, Santa Barbara, CA 93106}
\affiliation{Materials Department, University of California, Santa Barbara, CA 93106}

\author{Galan Moody}
\affiliation{Electrical and Computer Engineering Department, University of California, Santa Barbara, CA 93106}


\begin{abstract}
The development of manufacturable and scalable integrated nonlinear photonic materials is driving key technologies in diverse areas such as high-speed communications, signal processing, sensing, and quantum information. Here, we demonstrate a novel nonlinear platform---InGaP-on-insulator---optimized for visible-to-telecommunication wavelength $\chi^{\left(2\right)}$ nonlinear optical processes. In this work, we detail our 100-mm wafer-scale InGaP-on-insulator fabrication process realized via wafer bonding, optical lithography, and dry-etching techniques. The resulting wafers yield 1000s of components in each fabrication cycle, with initial designs that include chip-to-fiber couplers, 12.5-cm-long nested spiral waveguides, and arrays of microring resonators with free-spectral ranges spanning 400-900 GHz. We demonstrate intrinsic resonator quality factors as high as 324,000 (440,000) for single-resonance (split-resonance) modes near 1550 nm corresponding to 1.56 dB/cm (1.22 dB/cm) propagation loss. We analyze the loss versus waveguide width and resonator radius to establish the operating regime for optimal 775 nm – 1550 nm phase matching. By combining the high $\chi^{\left(2\right)}$ and $\chi^{\left(3\right)}$ optical nonlinearity of InGaP with wafer-scale fabrication and low propagation loss, these results open promising possibilities for entangled-photon, multi-photon, and squeezed light generation.
\end{abstract}

\maketitle

\begin{figure}[b!]
\centering
\includegraphics[width=0.75\columnwidth]{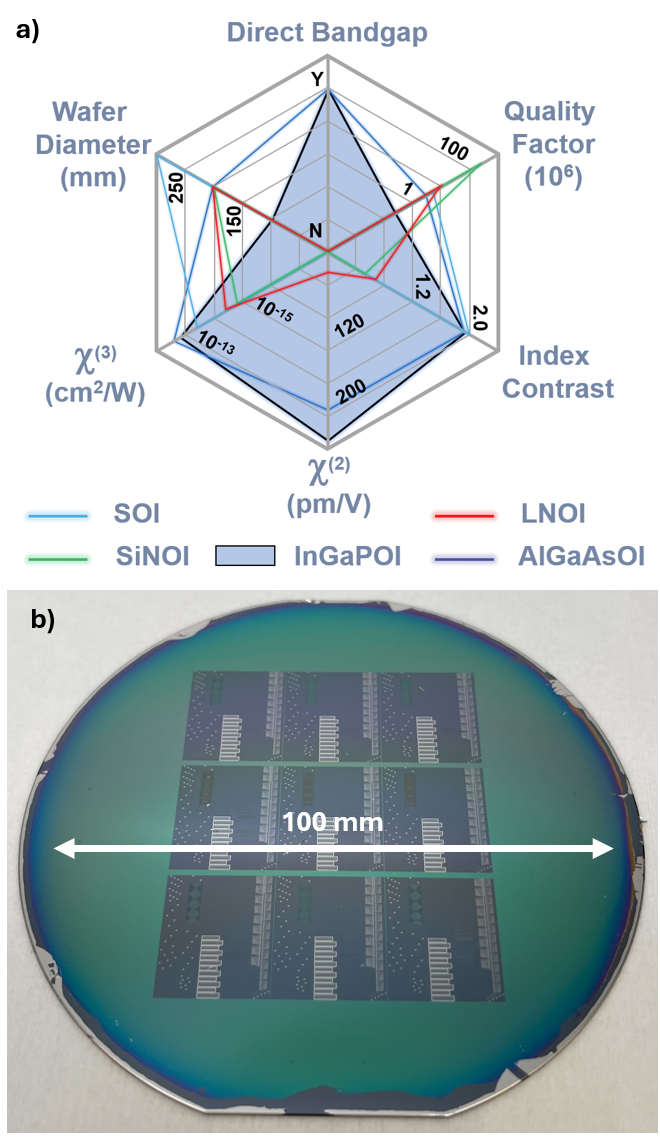}
\vspace{-10pt}
\caption{a) Comparison of material properties and capabilities for potential nonlinear quantum light source platforms. b) 100 mm InGaP-on-insulator (InGaPOI) wafer after full fabrication process.}
\label{fig:1}
\end{figure}

Chip-scale nonlinear quantum light sources that can efficiently generate bi-photon pairs at high rates and quality are a key component for optical quantum computing \cite{alexander2024manufacturable,arrazola2021quantum}, entanglement-based quantum key distribution \cite{steiner2023continuous,liu2022advances}, quantum time transfer \cite{lafler2023quantum}, and quantum sensing \cite{moody20222022}. Integrated photonic microring resonators comprised of nonlinear optical materials can generate entangled bi-photon pairs via $\chi^{(2)}$ spontaneous parametric down conversion (SPDC) or $\chi^{(3)}$ spontaneous four-wave mixing (SFWM) processes \cite{kues2019quantum,dutt2024nonlinear}, and they can be readily integrated with scalable chip-based photonic integrated circuits (PICs). Such sources have been demonstrated on many different material platforms, all of which face tradeoffs between optimal material properties and the constraints of existing fabrication technologies\cite{baboux2023nonlinear}, as shown in Fig. \ref{fig:1}(a). 

\begin{figure*}[t!]
\centering
\includegraphics[width=1\textwidth]{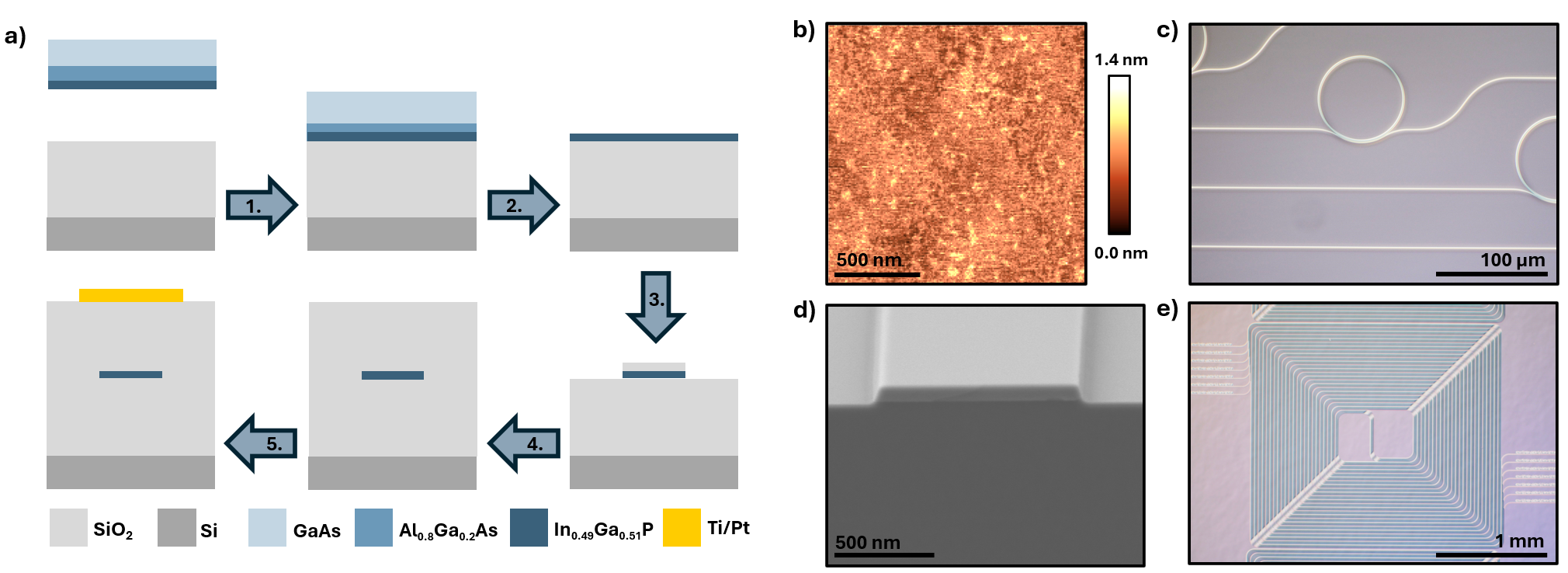}
    \vspace{-10pt}
\caption{a) InGaPOI process flow diagram. Step 1: Direct wafer-to-wafer bonding of InGaP to a thermally oxidized silicon base wafer. Step 2: Removal of GaAs growth substrate via wet etch. Step 3: Definition of waveguides in InGaP layer via deep-UV optical lithography and ICP etch. Step 4: ALD and PECVD deposition of oxide cladding layers. Step 5: Definition of Ti/Pt resistive heaters. b) Atomic Force Microscopy scan showing rms surface roughness of ~0.22 nm. c) Optical microscope image of ring resonators. d) Scanning electron microscope image of waveguide cross-section. e) Optical microscope image of nested spirals.}
\label{fig:2}
\end{figure*}

Silicon and silicon nitride photonics have already successfully scaled to mass production of low-loss PICs \cite{shekhar2024roadmapping, tran2022extending}, but they are at a disadvantage compared to other materials when it comes to optical nonlinearity \cite{moody2020chip}. Neither material has an intrinsic $\chi^{(2)}$ response; they also exhibit a weak $\chi^{(3)}$ nonlinearity (nitride) or suffer from two-photon and free-carrier absorption (silicon) at 1550 nm, which puts a fundamental limit on the performance of quantum light sources based on these platforms. Many different III-V materials including Gallium Arsenide (GaAs), Aluminum Gallium Arsenide (AlGaAs), Aluminum Nitride (AlN), and Indium Phosphide (InP) are attractive for this purpose because of their strong $\chi^{(2)}$ and $\chi^{(3)}$ responses, their large index contrast, and, with advances in fabrication techniques, they have enabled bright entangled-photon pair sources \cite{mobini2022algaas,chang2022csoi,guo2017parametric,kumar2019entangled}. SFWM from AlGaAs-on-insulator resonators, for example, has enabled $>$~20 GHz/mW$^2$ pair generation rates, coincidence-to-accidental (CAR) ratio $>$~3,000, up to $10^6$ detected coincidences per second, better than $99\%$ heralded single-photon purity, and entanglement visibility $>99\%$ \cite{Steiner2021,castro2022expanding}. These chip-scale sources now produce entangled-photon pairs with rates that are competitive with the best solid-state single-photon emitters and bulk nonlinear optical systems, but with the inherent scalability, compactness, and efficiency enabled by room-temperature PIC devices \cite{moody2020chip}. 

Compared to SFWM, for materials with broken inversion symmetry, the stronger $\chi^{(2)}$ response can more efficiently generate photon pairs through spontaneous parametric down-conversion (SPDC) \cite{appas2022nonlinear,placke2024telecom,appas2023broadband}. One promising $\chi^{(2)}$ material is indium gallium phosphide (In$_{0.49}$Ga$_{0.51}$P) lattice-matched to GaAs---a III-V semiconductor with underdeveloped potential. In addition to its high $\chi^{(2)}$ and $\chi^{(3)}$ nonlinearities ($\chi^{(2)}$ $\approx$ 220 pm/V, 1.5 times larger than AlGaAs and 10 times larger than lithium niobate), InGaP has a relatively wide bandgap of 1.9 eV (645 nm) compared to many other III-V materials developed for quantum PICs\cite{Ueno1997}. As a result, InGaP devices have already demonstrated efficient pair generation using SFWM and SPDC in suspended microcavities fabricated with electron-beam lithography \cite{zhao2022ingap,akin2024ingap,chopin2023ultra}. If the challenge of scalable fabrication of high quality devices can be overcome, InGaP has the potential to become a leading platform for integrated quantum photonics. This challenge is nontrivial, as shown by years of work on more established platforms including silicon, silicon nitride, and AlGaAs. To that end, we have developed an InGaP-on-insulator (InGaPOI) process that can meet the requirements of scale and compatibility for integration. We present wafer-scale fabrication of InGaPOI microring resonators and 12.5-cm-long spirals with propagation losses as low as 1.22 dB/cm and intrinsic quality factors greater than 4 $\times$ 10$^5$ at 1550 nm, demonstrated for devices across a 100 mm wafer shown in Fig. \ref{fig:1}(b). To the best of our knowledge, these results are the first 100 mm wafer-scale fabrication process demonstrated for InGaP-on-insulator which can be immediately scaled to 200 mm production with our existing wafer bonding capabilities, opening exciting prospects for manufacturing highly nonlinear quantum photonic wafers.

The InGaPOI fabrication process begins with a wafer-scale, low-temperature plasma-activated bonding process which leverages expertise in substrate-transferred epitaxial structures for the development of compound semiconductor on insulator devices \cite{chang2022csoi} and high-performance precision laser optics \cite{cole2023substrate}. The InGaP epitaxial wafer (MOCVD-grown by Twenty-One Semiconductors) was inspected to ensure its suitability for bonding. Atomic force and scanning optical microscopy were used to confirm RMS surface microroughness $<$ 1 nm and micrometer-scale defect density $<<1000$ cm$^{-2}$. The wafer was then thoroughly rinsed in a spin-cleaning system with a combination of solvents (acetone and isopropanol), a nonionic surfactant, and deionized water. After cleaning, the wafer was re-inspected to ensure no particles or remnant organic contaminants remain on its surface. 

Directly before bonding, an oxygen plasma activation process was carried out on both the InGaP and thermal-oxide-on-silicon substrate wafers. After a final inspection, the two wafers were loaded into a commercial bonding tool (EVG 520 IS semi-automated wafer bonding system). The bond chamber was evacuated to a pressure $< 1\times10^{-4}$ mbar and the wafers were mechanically pressed together at room temperature with 9 kN of force. A 12-hour-long anneal at 150 \textdegree C was then performed external to the bonder to enhance the interfacial energy of the contacted wafers. 

\begin{figure*}[t!]
\centering
\includegraphics[width=1\textwidth]{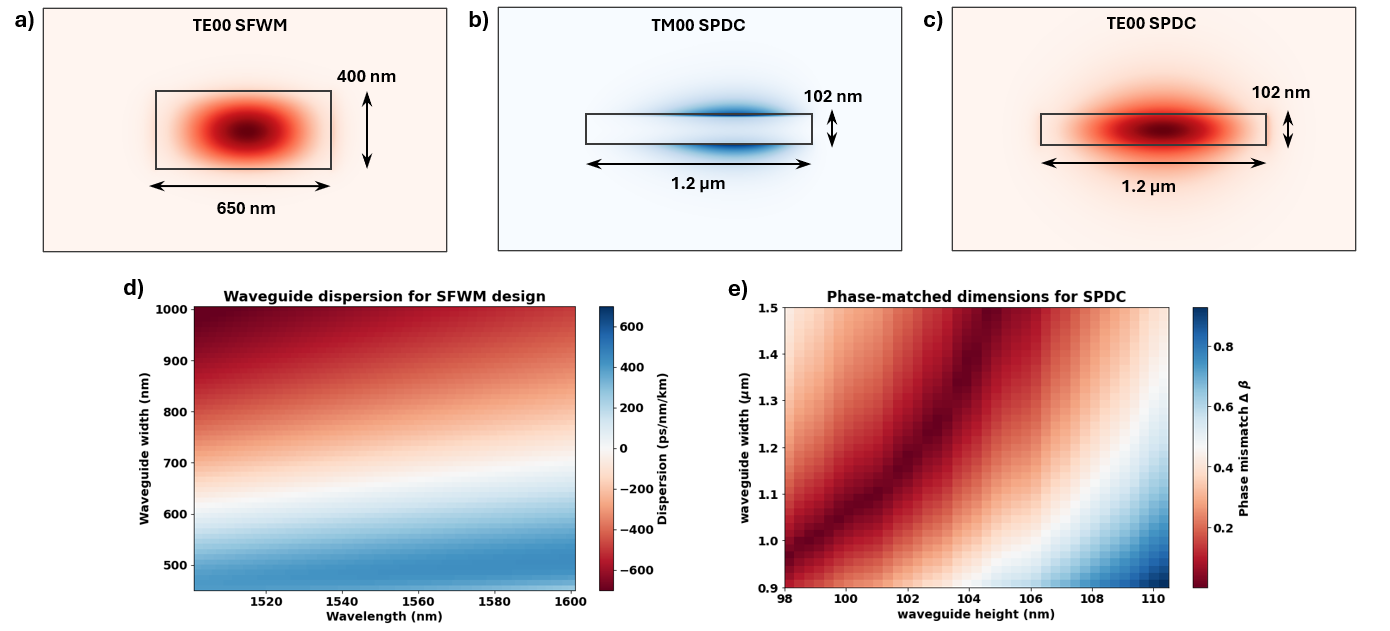}
    \vspace{-10pt}
\caption{a)  Modal cross-section for the phase-matched SFWM 1550 nm TE$_{00}$ mode. b) Modal cross-section for the phase-matched SPDC TM$_{00}$ mode. c) Modal cross-section for optimized phase-matched SPDC 1550 nm TE$_{00}$. d) Waveguide dispersion near 1550 nm for SFWM designs of InGaP with 400nm thickness. e) Frequency mismatch versus waveguide width for 102-nm thick InGaP  775 nm – 1550 nm SPDC designs.}
\label{fig:3}
\end{figure*}

Removal of the GaAs substrate was performed using a NH$_4$OH:H$_2$O$_2$ wet etch without any prior mechanical lapping of the GaAs wafer. The etch terminates on the Al$_{0.8}$Ga$_{0.2}$As etch stop layer, which is then selectively removed with a dilute HF-based wet chemical etch. The surface of the 102-nm-thick InGaP waveguiding layer was thoroughly cleaned and inspected to verify surface quality and defect density before further processing. For waveguide feature definition, a 90-nm thick SiO$_2$ hard mask was deposited using atomic layer deposition (ALD). An anti-reflection coating and photoresist (DUV42P6 and UV0.8-6) were spun-on and then patterned via deep-UV optical lithography with an ASML stepper tool. After development, a thermal reflow process was used to smooth the sidewall roughness in the photoresist that may transfer to the hard mask and then waveguide profile during etching \cite{xie2020ultrahigh}. The reflow process must be carefully calibrated to maximize improvement in sidewall roughness while preserving feature size and uniformity across the wafer. 

After the reflow process, an inductively coupled plasma (ICP) etch was used to define the features in the hard mask, followed by thorough cleaning to remove resist and AR coating residue. A BCl$_3$/Cl$_2$/N$_2$ ICP etch was then used to define the waveguides in the InGaP. Next 30 nm of ALD SiO$_2$ and 1.5 $\mu$m of PECVD SiO$_2$ were deposited as waveguide cladding, and Ti/Pt resistive heaters were patterned on top of the cladding above specific regions of devices for thermo-optic phase tuning. Finally, the wafer underwent a facet-etching process and was diced into 45 separate die for screening and characterization.

Both SFWM and SPDC processes are possible with InGaP-on-insulator devices; however, optimal designs for each process require very different waveguide cross-sections. For any nonlinear process, conservation of energy, phase matching,  high confinement of the optical mode within the waveguide, and substantial mode overlap are important for optimizing the efficiency. For SFWM of the fundamental transverse electric (TE) mode, a near-zero dispersion waveguide design allows for generation of entangled pairs spanning a larger bandwidth for waveguides and a larger range of frequency bins for a single resonator. Using the dispersion curves from Tanaka \textit{et al.}\cite{Tanaka1986}, we calculate the waveguide dispersion for the fundamental TE mode at 1550 nm for various SFWM designs and find an optimal cross-section of 400 nm $\times$ 650 nm (Fig. \ref{fig:3}(a) and Fig. \ref{fig:3}(d)). 

\begin{figure}[!b]
\centering
\includegraphics[width=1\columnwidth]{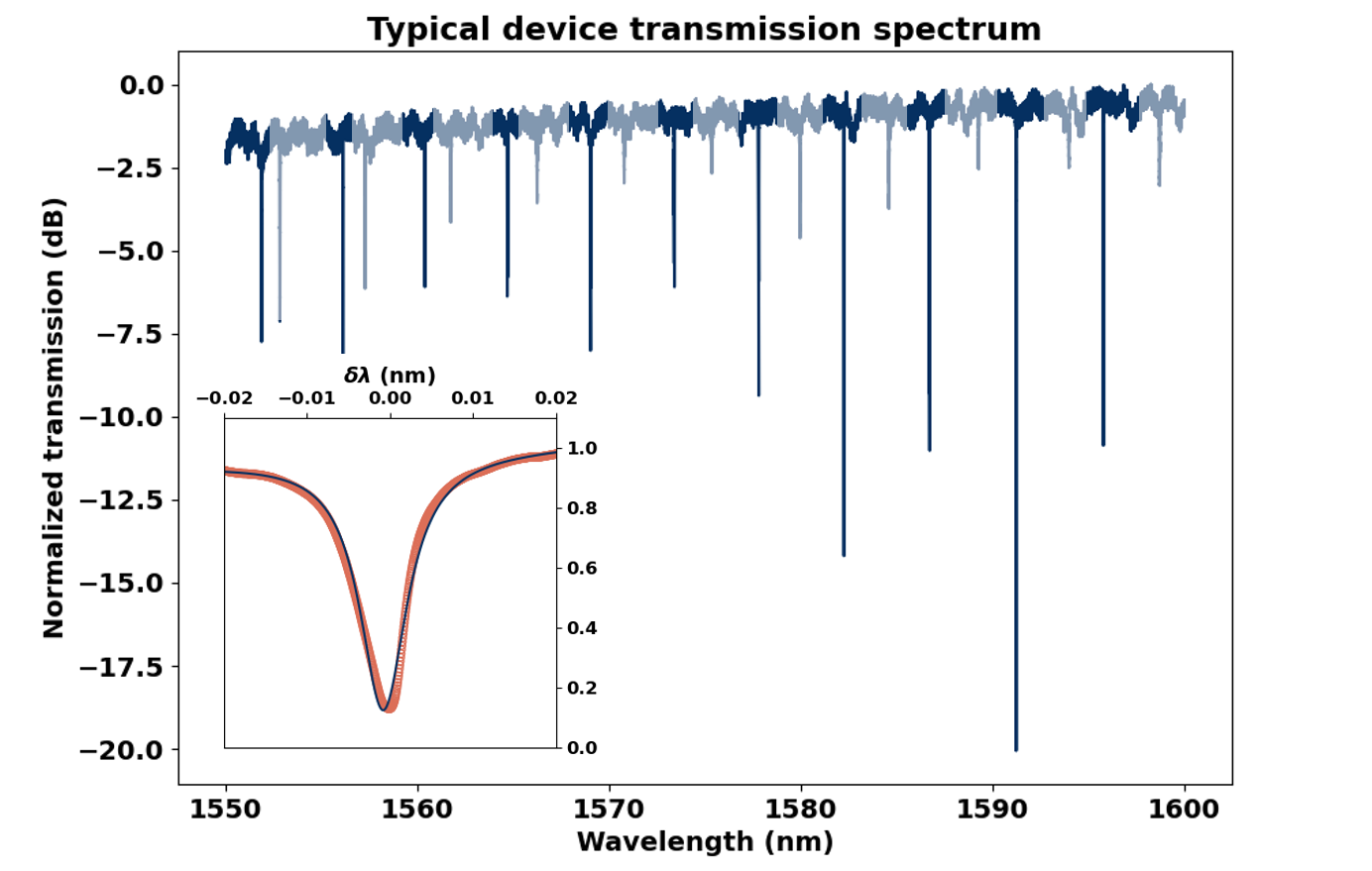}
\vspace{-10pt}
\caption{Typical ring resonator transmission spectrum over 1550-1600 nm. Free-spectral range analysis was used to identified the 1550 nm TE$_{00}$ modes as the higher extinction mode family.  Inset) Lorentzian fit for resonance with $Q_L$= 213,000, $Q_i$= 342,000, $\alpha$ = 1.56 dB/cm}
\label{fig:4}
\end{figure}

\begin{figure*}[!t]
\centering
\includegraphics[width=1\textwidth]{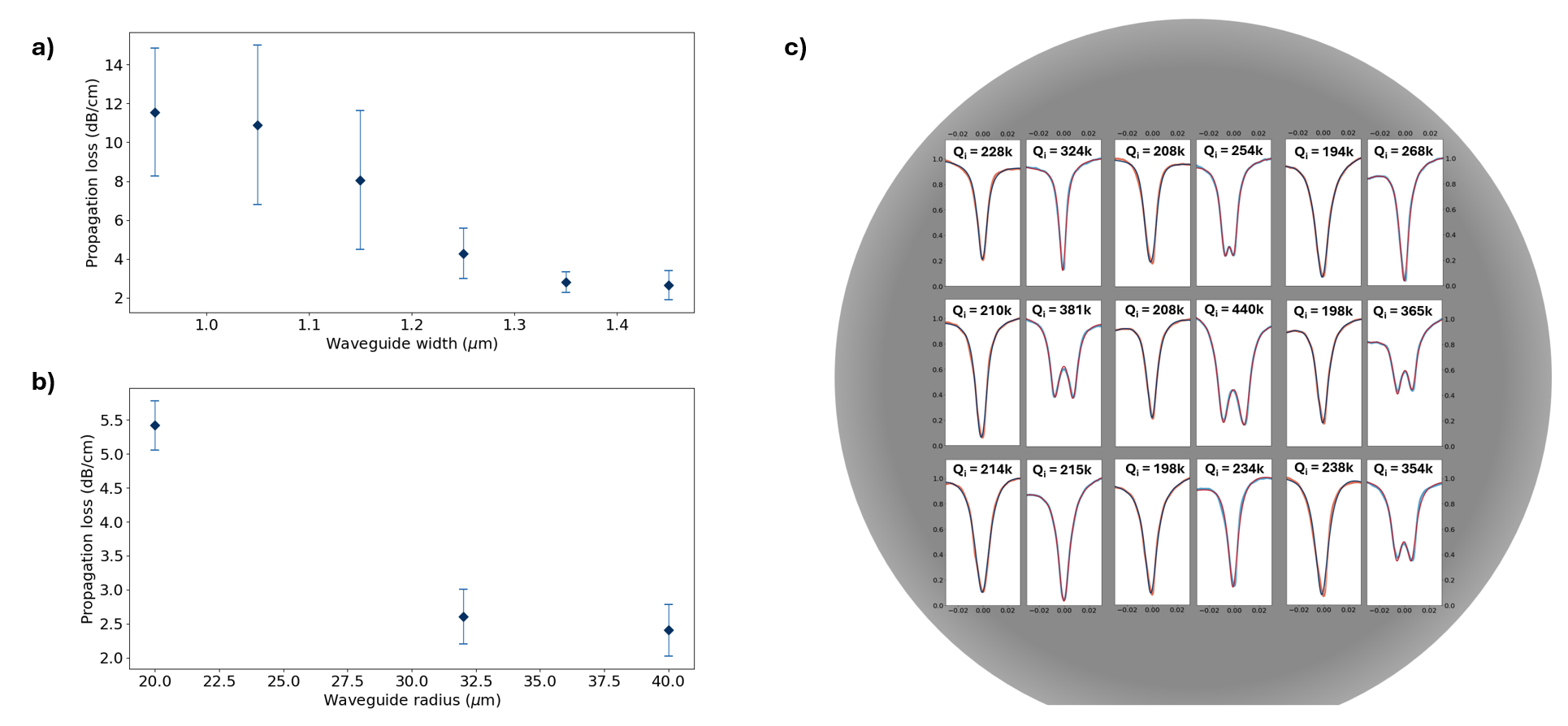}
    \vspace{-10pt}
\caption{a) Propagation loss versus ring waveguide width for rings with radius = 32 $\mu$m. b) Propagation loss versus radius for rings with 1.45 $\mu$m width. For a) and b), values are each averaged across identical device designs from at least three die, looking at device designs with closest to critical coupling. c) Highest intrinsic quality factors for two devices from each of nine die across the wafer.  On left of each die panel, identical device design with radius = 32 $\mu$m and width = 1.45 $\mu$m. On right, device with any design parameters with highest intrinsic quality factor. Devices included have radii of 32-40 $\mu$m and widths of 1.35-1.55 $\mu$m. Resonances are fit using the split resonance transfer function described in the text with a sinusoidal background correction to account for minor facet reflections.}
\label{fig:5}
\end{figure*}

Because of the strong dispersion of InGaP (refractive index $n$ = 3.41 at 775 nm and $n$ = 3.12 at 1550 nm), quasi-phase matching for SPDC can only be obtained by using high-aspect-ratio cross sections as shown in Fig. \ref{fig:3}(b) and Fig. \ref{fig:3}(c) for the fundamental TM mode at 775 nm and fundamental TE mode as 1550 nm, respectively. Considering conservation of energy ($2\omega_{TE,1550} = \omega_{TM,775}$) and the phase-matching condition $|2m_{TE,1550} - m_{TM,775}| = 2$, where $\omega$ is the angular frequency of the indicated mode and $m$ is the azimuthal number of the mode in a resonator, we show that in the resonator, the phase-matched waveguide width varies sharply with waveguide height, presenting additional considerations for designing resonators for SPDC. For a waveguide height of 102 nm, the ideal width is $\sim$ 1.2 $\mu$m, although a 2 nm change in waveguide height requires a $\sim$100 nm change in the waveguide width to achieve phase matching.  

With these considerations, we fabricated a full wafer (Fig. \ref{fig:1}(b)) that includes nine nominally identical dies (with five sub-dies on each for a total of 45 chips) of ring resonator and spiral waveguide designs with various parameter sweeps. After die singulation, individual chips are tested in a custom-built PIC testbed with fiber-to-chip end coupling using lensed polarization-maintaining fibers, chip temperature stabilization at 20 \textdegree C, electrical probes for thermo-optic tuning, and a microscope for probe and fiber alignment. For each device, light is coupled on and off chip via waveguide tapers. For resonator devices, pulley coupler designs are used for waveguide bus-to-ring power transfer. The parameters swept include ring radius, ring waveguide width, and pulley coupler bus waveguide width and angle. For the pulley coupler designs, the coupling gap was fixed at 300 nm and the angle is swept to control the coupling. Transmission spectra for each device were measured using a continuously tunable laser scanning across 1530-1600 nm. Light coupled off the chip was measured with a fast photodiode and oscilloscope with a separate fiber-based Mach Zehnder interferometer for wavelength calibration. In this work, we focus on characterization of the performance of devices near 1550 nm.

A representative linear transmission spectrum is shown in Fig. \ref{fig:4} for a resonator with 32 $\mu$m radius and 1.35 $\mu$m width. For SPDC ring designs, the dimensions required for quasi-phase-matching result in resonators that can support higher-order TE modes in the telecom wavelength range. For devices with multiple mode families, the target fundamental TE mode was identified and isolated by analyzing the free spectral range of each mode family. Resonances associated with the fundamental TE mode are highlighted in dark blue in the spectrum shown in Fig. \ref{fig:4}. Each resonance is fit using an analytical model developed by Moresco \textit{et al.} that takes into account the interplay between standing-wave modes due to back-scattering that leads to resonance splitting in some of the devices, and the traveling ring modes in all of the devices \cite{moresco2013method}. From the fits, we can determine the loaded quality factor $Q_L$, intrinsic quality factor $Q_i$ that is inversely proportional to the waveguide propagation loss $\alpha$, and the reflection coefficient $\rho$ that determines the correlation length and root-mean-square value of the surface roughness that leads to the back-scatter. For modes without resonance splitting, fits from the model agree with a Lorentzian fit to within $\sim1\%$. The inset to Fig. \ref{fig:4} shows that $Q_i$ up to 324,000 is measured from this particular device corresponding to a propagation loss of 1.56 dB/cm.

Parameter sweeps of ring designs can be used to extract trends in propagation loss as a function of ring radius and ring waveguide width, both of which help assess the potential performance of InGaPOI devices for nonlinear frequency conversion and quantum light generation. Maintaining low propagation loss at small ring radii enables smaller mode volumes and therefore higher conversion efficiency. Figure \ref{fig:5}(a) shows the average propagation loss measured from three die across the wafer for increasing microring resonator radius from 20 to 40 $\mu$m. Nominally identical resonators from each die with 1.45 $\mu$m waveguide widths are chosen for device designs closest to critical coupling to the bus waveguide. With increasing radius the average propagation loss decreases from $\sim5.4$ dB/cm to $\sim2.4$ dB/cm. From this curve, we estimate that for radii below $\sim20-25$ $\mu$m, scattering from the resonator due to greater mode overlap with the sidewall begins to dominate over material absorption loss and governs the loaded quality factor $Q_L$. 

We next examined the average propagation loss for different resonator waveguide widths, shown in Fig. \ref{fig:5}(b). For increasing width, the loss decreases down to $\sim2.5$ dB/cm as a result of less mode overlap with the etched waveguide sidewalls. For both data from the radius and width sweeps in Fig. \ref{fig:5}, error bars are shown as the standard deviation from the mean of at least three different nominally identical devices. For smaller radius and width, larger device-to-device variation is observed due to the larger mode interaction with the sidewalls. In these designs, fabrication process variations have a greater impact on scattering and thus the measured $Q$. Figure \ref{fig:5}(b) illustrates that low propagation loss over a range of ring waveguide widths ensures that the high device performance demonstrated here reflects that of device designs that also meet the phase-matching requirements for this waveguide thickness.

Finally, we show in Fig. \ref{fig:5}(c) linear transmission plots from the highest-$Q$ devices from each die across the wafer, ranging from 194,000 to 440,000 (not necessarily similar resonator FSR or waveguide widths). We see that for some of the highest-$Q$ resonators, the resonances are split due to backscattering, which is more apparent with increasing $Q$ as expected from the model by Moresco \textit{et al.} discussed previously. In Table 1, we compare the linear propagation loss determined from these measurements to those reported for other $\chi^{\left(2\right)}$ photonic platforms that have been developed specifically for nonlinear optical conversion and pair generation processes. 

Notably, while the loss from many other platforms have reached sub-dB/cm, excluding AlGaAsOI they are all fabricated using electron-beam lithography (EBL), compared to 100-mm wafer scale fabrication using deep-UV photolithography in our study. The 1.22 dB/cm propagation loss at 1550 nm reported here is the lowest amongst the most recently proposed heterogeneously integrated nonlinear platforms (ScAlNOI and GaNOI), with strong outlook for improvements in device performance. Surface passivation of InGaP with an Al$_2$O$_3$ layer has been shown to provide a $3\times$ improvement in the intrinsic quality factor at telecom wavelengths\cite{akin2024ingap}. Use of deuterated SiO$_2$ has been shown to reduce material absorption from the cladding layer by $\sim7\times$ at 1550 nm and has the potential to provide performance improvement across the visible wavelength range \cite{Jin2020}. 

\begin{table}[t!]
    \caption{A comparison of the 1550 nm linear propagation loss for different $\chi^{\left(2\right)}$ integrated photonic platforms developed for nonlinear frequency conversion and quantum light generation. The GaNOI loss is estimated from the published data; the other data are reported from the listed citations. DUV-PL: deep-UV photolithography. EBL: electron-beam lithography.}
    \label{tab:Table1}
    \begin{ruledtabular}
    \centering
    \begin{tabular}{|c|c|c|}
           Nonlinear & Propagation Loss & Fabrication \\
           Material & @ 1550 nm & Method \\
         \hline
         InGaPOI (this work) & 1.22 dB/cm& DUV-PL, 100-mm wafer \\
         \hline
         Suspended InGaP \cite{akin2024ingap} & 0.8 dB/cm & EBL \\
         \hline
         AlGaAsOI \cite{xie2020ultrahigh} & 0.2 dB/cm & DUV-PL, 100-mm wafer \\
         \hline
         LNOI \cite{luo2018highly} & 0.5 dB/cm & EBL \\
         \hline
         GaNOI \cite{zeng2024quantum} & $\sim$2 dB/cm & EBL \\
        \hline
         ScAlNOI \cite{friedman2024measured} & 1.6 dB/cm & EBL \\
        \hline
         SiC \cite{guidry2020optical} & 0.38 dB/cm & EBL \\
    \end{tabular}
    \end{ruledtabular}
\end{table}

In conclusion, we presented a new InGaP-on-insulator 100-mm wafer-scale fabrication process with 200 nm feature sizes using direct wafer bonding and substrate removal, deep-UV photolithography, and dry etching methods. Each wafer produces 1000s of photonic components with waveguide propagation loss down to 1.22 dB/cm and 1550 nm resonator quality factors up to 440,000. Numerical and experimental characterization of the loss and dispersion with waveguide cross section establishes the optimal regime for SFWM and SPDC nonlinear processes. Considering the large $\chi^{\left(2\right)}$ and $\chi^{\left(3\right)}$ nonlinearities, wide bandgap (1.9 eV) and high refractive index ($>3$), these results establish InGaP as a promising candidate for a wide range of nonlinear photonic applications, including second harmonic generation, quantum frequency conversion, broadband entangled-pair generation, and squeezed light generation.

\begin{acknowledgments}
\noindent A portion of this work was performed in the UCSB Nanofabrication Facility, an open access laboratory. We thank Ifeanyi Achu, Navin Lingaraju, Joseph Christesen, and Cale Gentry from SRI International for technical input and assistance in device modeling and designs. Wafers were obtained from Twenty-One Semiconductors GmbH. This work was supported by the NSF Quantum Foundry through the Q-AMASE-i Program (Grant No. DMR-1906325), the NSF CAREER Program (Grant No. 2045246), and the Air Force Office of Scientific Research (Grant No. FA9550-23-1-0525). L.T. acknowledges support from the NSF Graduate Research Fellowship Program.
\end{acknowledgments}

\section*{Data Availability Statement}

\noindent The data that support the findings of this study are available from the corresponding author upon reasonable request.

\section*{Author Declarations}

\noindent The authors have no conflicts to disclose.

\bibliography{moodybib}

\end{document}